\documentclass[english]{article}
\usepackage[T1]{fontenc}
\usepackage[latin9]{inputenc}
\usepackage[letterpaper]{geometry}
\makeatletter

\providecommand{\tabularnewline}{\\}

\makeatother

\usepackage{babel}

\begin{document}
\textbf{\large $N$-qubit Quantum Teleportation, Information Splitting
and Superdense Coding }{\large \par}

\textbf{\large ~~~~~~~~~~~~~~~~~~~~~~~~~through
the composite GHZ-Bell Channel}{\large \par}

~

~~~~~~~~~~~~~~~~~~~~~~~~~~~~~~~~~~~~~~~~~~
Debashis Saha$^{\dagger}$ and Prasanta K. Panigrahi$^{*}$

\textit{~~~~~~~~~~~Indian Institute of Science Education
and Research (IISER) Kolkata, Mohanpur Campus}

~~~~~~~~~~~~~~~~~~BCKV Campus Main Office, Mohanpur
- 741252, Nadia, West Bengal, India

~

~
\begin{abstract}
We introduce a general odd qubit entangled system composed of GHZ
and Bell pairs and explicate its usefulness for quantum teleportation,
information splitting and superdense coding. After demonstrating the
superdense coding protocol on the five qubit system, we prove that
`$2N+1$' classical bits can be sent by sending `$N+1$' quantum bits
using this channel. It is found that the five-qubit system is also
ideal for arbitrary one qubit and two qubit teleportation and quantum
information splitting (QIS). For the single qubit QIS, three different
protocols are feasible, whereas for the two qubit QIS, only one protocol
exists. Protocols for the arbitrary $N$-qubit state teleportation
and quantum information splitting are then illustrated.

~~

~~

~~
\end{abstract}

\subsection*{~~~~~~~~~~~~~~~~~~~~~~~~~~~~~~~~~~~~~~~~{\normalsize ~
~I. INTRODUCTION }}

Entanglement is the fascinating aspect of quantum mechanics, due to
which non-intuitive quantum correlations can exist between two or
more particles {[}1{]}. Using entanglement, one can perform different
types of quantum tasks like teleportation {[}2{]}, superdense coding
{[}3{]}, secret sharing {[}4,5{]}, quantum cryptography, one-way quantum
computation {[}6{]} etc. Teleportation is a unique quantum procedure
that allows one party to send a quantum state to another, without
the state being physically transmitted. Teleportation of arbitrary
single qubit, through an entangled channel of Einstein-Podolsky-Rosen
{[}EPR{]} pair between the sender and receiver, was fi{}rst demonstrated
by Bennett et al., {[}2{]}. Experimentally it has been achieved using
different quantum systems, inside and outside laboratory conditions
{[}7-11{]}. Another related protocol is superdense coding, which allows
one to send two classical bits of information encoded in a single
qubit using a bipartite entangled channel {[}3{]}. `Quantum information
splitting' is a quantum state sharing protocol, where sharing of quantum
information can be done between a group of parties, such that none
of them can reconstruct the unknown information, without knowing the
others' measurement outcomes. This was first demonstrated for a single
qubit state, using three qubit Greenberger-Horne-Zeilinger (GHZ) state,
with three agents having one qubit each {[}12{]}. Experimental realization
of QIS has been achieved using several states {[}13-15{]} and single
photon sources {[}14{]}.

A number of multipartite entangled states like the GHZ states {[}16{]},
modified W states {[}17,18{]}, magnon state {[}19{]}, mirror state
{[}20{]} and the cluster states {[}21,22{]} have been introduced and
exploited for carrying out several quantum tasks. Recently, a genuinely
entangled five-qubit state {[}23{]} and a six-qubit state {[}24{]}
have been used for successfully carrying out many quantum communication
protocols {[}25,26{]}. Superdense coding has been done using combined
Bell states, where `$2N$' bits of classical information is transmitted
via `$N$' qubits {[}27{]}. $N$-qubit teleportation has been discussed
in several cases using different entangled quantum channels {[}28,29{]}.
Also, a number of QIS protocols have been recently implemented {[}30-32{]}.

In this paper, we demonstrate the utility of a quantum channel, composed
of GHZ and Bell pairs for various quantum communication protocols.
It has been shown that, some of the Bell states are decoherence free
under certain environment {[}33{]}. Being the superposition of two
terms, the GHZ state can be less prone to decoherence. This fact gives
our primary motivation for studying such state. Before explicating
those protocols, we investigate the nature of entanglement between
the subparties of the five-qubit composite GHZ-Bell state:

$|\zeta\rangle=\frac{1}{\sqrt{2}}(|000\rangle+|111\rangle)\frac{1}{\sqrt{2}}(|00\rangle+|11\rangle)=\frac{1}{2}(|00000\rangle+|00011\rangle+|11100\rangle+|11111\rangle)$.~~~~~~~~~~~~~~~~~~~~~~~~~~~~~~~~~
(1)

~

~

---------------

~

{\footnotesize $*$ electronic address : pprasanta@iiserkol.ac.in}{\footnotesize \par}

{\footnotesize $\dagger$ electronic address : debashis7112@iiserkol.ac.in}{\footnotesize \par}

Consider $|\zeta\rangle$ as 
$\frac{1}{2}(|0_{A}0_{A}0_{B}\rangle+|1_{A}1_{A}1_{B}\rangle)(|0_{A}0_{B}\rangle+|1_{A}1_{B}\rangle)$,
where Alice has three qubits and Bob has two qubits. The reduced density
matrix obtained by partial tracing over the sub-system `A' is given
by,

$\rho_{B}=Tr_{A}|\zeta\rangle\langle\zeta|=\frac{1}{4}(|00\rangle_{B}\langle00|_{B}+|01\rangle_{B}\langle01|_{B}+|10\rangle_{B}\langle10|_{B}+|11\rangle_{B}\langle11|_{B})=\frac{1}{4}\left(\begin{array}{cccc}
1 & 0 & 0 & 0\\
0 & 1 & 0 & 0\\
0 & 0 & 1 & 0\\
0 & 0 & 0 & 1\end{array}\right)$ which is maximally mixed. For another inequivalent distribution,
$|\zeta\rangle$= $\frac{1}{2}(|0_{A}0_{A}0_{A}\rangle+|1_{A}1_{A}1_{A}\rangle)(|0_{A}0_{B}\rangle+|1_{A}1_{B}\rangle)$,
the reduced density matrix is given by, 
$\rho_{B}=Tr_{A}|\zeta\rangle\langle\zeta|=\frac{1}{4}(|0\rangle_{B}\langle0|_{B}+|1\rangle_{B}\langle1|_{B})=\frac{1}{4}\left(\begin{array}{cc}
1 & 0\\
0 & 1\end{array}\right)$. Hence, the state $|\zeta\rangle$ has the maximum possible entanglement
between two subsystem satisfying the general condition for teleportation
{[}34{]}. In the following section, we first show the superdense coding
of five bits of information using this state and subsequently generalize
it to `$2N+1$' bits of information. In the section III, we explicate
arbitrary one, two and $N$-qubit state teleportation using composite
GHZ-Bell measurement. The next section deals with QIS of one and two-qubit
states using different protocols and show that this channel can also
be used for $N$-qubit QIS. We then conclude in the last section,
with several directions for future work. We now procced to explicate
various protocols starting with superdense coding.

~~

~

\subsection*{~~~~~~~~~~~~~~~~~~~~~~~~~~~~~~~~~~~~~~{\normalsize II.
SUPERDENSE CODING }}

We start with the protocol of superdense coding of five classical
bit, through the above mentioned GHZ-Bell channel: 
$|\zeta\rangle=\frac{1}{2}(|0_{A}0_{A}0_{B}\rangle+|1_{A}1_{A}1_{B}\rangle)(|0_{A}0_{B}\rangle+|1_{A}1_{B}\rangle)$.{\small{}
}Alice and Bob know the correspondence of quantum state and classical
bit string. Alice can transform the state $|\zeta\rangle$ into 32
entangled orthonormal states by applying Paulli matrices for encoding
classical bits. Depending on which bit string Alice wants to send
to Bob, she has to perform an appropriate unitary operation on her
qubits. Alice then sends her qubits to Bob and by making a measurement
using the orthonormal states given in Table I, Bob successfully gets
the five bits of classical information. In Table I, we have shown
that all the orthonormal states can be written as product of the following
states:

$|\xi^{\pm}\rangle=\frac{1}{\sqrt{2}}(|000\rangle\pm|111\rangle)$,
$|\chi^{\pm}\rangle=\frac{1}{\sqrt{2}}(|011\rangle\pm|100\rangle)$,
$|\vartheta^{\pm}\rangle=\frac{1}{\sqrt{2}}(|010\rangle\pm|101\rangle)$,
$|\theta^{\pm}\rangle=\frac{1}{\sqrt{2}}(|001\rangle\pm|110\rangle)$,

$|\psi^{\pm}\rangle=\frac{1}{\sqrt{2}}(|00\rangle\pm|11\rangle)$,
$|\phi^{\pm}\rangle=\frac{1}{\sqrt{2}}(|01\rangle\pm|10\rangle)$.~~~~~~~~~~~~~~~~~~~~~~~~~~~~~~~~~~~~~~~~~~~~~~~~~~~~~~~~~~~~~~~~~~~~~~~~~~~~~~~~~~~~~~~~~~~~~~~~~~(2)

Hereforth, we use $I$ as the identity matrix, $X$ as $\sigma_{x}$,
$Z$ as $\sigma_{z}$ and $U_{i}$ denotes the operator, where $U$
acts on the $i$-th qubit from the left hand side.

\subsubsection*{Generalized superdense coding of `$2N+1$' bits of information}

This superdense coding can be generalized to $(2N+1)$-qubit channel.
For transmiting `$2N+1$' bits of information, Alice and Bob share
a multipartite entangled state, which is the product of one $|\xi^{+}\rangle$
and ($N-1$) number of $|\psi^{+}\rangle$ states such that Alice
possesses `$N+1$' qubits and Bob possesses `$N$' qubits:

$|\zeta_{0}\rangle=|\xi^{+}\rangle_{AB}|\psi^{+}\rangle_{AB}........|\psi^{+}\rangle_{AB}$
,~~~~~~~~~~~~~~~~~~~~~~~~~~~~~~~~~~~~~~~~~~~~~~~~~~~~~~~~~~~~~~~~~~~~~~~~~~~~~~~~~~~~~~~~~~~~~~~~(3)
where~~~~$|\xi^{+}\rangle_{AB}=\frac{1}{\sqrt{2}}(|0_{A}0_{A}0_{B}\rangle+|1_{A}1_{A}1_{B}\rangle)$
and $|\psi^{+}\rangle_{AB}=\frac{1}{\sqrt{2}}(|0_{A}0_{B}\rangle+|1_{A}1_{B}\rangle)$.
The capacity of superdense coding of this channel is $2N+1$, satisfying
`Holevo bound\textquoteright{} {[}35{]}. Alice wants to send a classical
bit string, which is $a_{2N+1}a_{2N}.......a_{2}a_{1}$ ($a_{i}=0$
or $1$) and the corresponding quantum state is $|\zeta_{j}\rangle$,
where `$j$' is the decimal representation of $a_{2N+1}a_{2N}.......a_{2}a_{1}$
. Here, `$j$' can take value from 0 to $2^{2N+1}-1$. The unitary
operations performed by Alice to obtain the quantum state $|\zeta_{j}\rangle$
is given by,

$|\zeta_{j}\rangle=\otimes_{k=1}^{N}(Z_{2k})^{a_{k}}\otimes_{l=1}^{N+1}(X_{|2l-2|})^{a_{l+N}}|\zeta_{0}\rangle$.
~~~~~~~~~~~~~~~~~~~~~~~~~~~~~~~~~~~~~~~~~~~~~~~~~~~~~~~~~~~~~~~~~~~~~~~~~~~~~~~~~~~~~~~~~~(4)
It is noted that, $I$ is referred as $Z^{0}$ or $X^{0}$. All $|\zeta_{j}\rangle$'s
are mutually orthonormal and can be written as products of $|\xi^{\pm}\rangle_{AB},|\chi^{\pm}\rangle_{AB},$
$|\vartheta^{\pm}\rangle_{AB},|\theta^{\pm}\rangle_{AB},|\psi^{\pm}\rangle_{AB},|\phi^{\pm}\rangle_{AB}$
states. The subscript $AB$ denotes that the last particle of that
state belongs to Bob and others belong to Alice. After applying the
unitary operation, Alice has to send her `$N+1$' qubits to Bob and
he will be able to get `$2N+1$' bits of information by making a measurement
in the $\left\{ |\zeta_{j}\rangle\right\} $ basis, which satisfy
the superdense protocol perfectly.

\subsubsection*{{\footnotesize TABLE I: The unitary operations performed by Alice
on her qubits and the resulting 32 orthonormal states $|\zeta_{j}\rangle$
($j=0,1,..,31$) with their corresponding classical bit strings. }}

\begin{tabular}{|c|c|c|c|}
\hline
{\footnotesize Classical bit} & {\footnotesize Unitary operation} & {\footnotesize State obtained} & {\footnotesize Decomposition 
of the state}\tabularnewline
\hline
\hline
{\footnotesize 00000} & {\footnotesize $I$} & {\footnotesize 
$|\zeta_{0}\rangle=\frac{1}{2}(|00000\rangle+|00011\rangle+|11100\rangle+|11111\rangle)$} & {\footnotesize 
$|\xi^{+}\rangle_{AB}|\psi^{+}\rangle_{AB}$}\tabularnewline
\hline
{\footnotesize 00001} & {\footnotesize $Z_{2}$} & {\footnotesize 
$|\zeta_{1}\rangle=\frac{1}{2}(|00000\rangle+|00011\rangle-|11100\rangle-|11111\rangle)$} & {\footnotesize 
$|\xi^{-}\rangle_{AB}|\psi^{+}\rangle_{AB}$}\tabularnewline
\hline
{\footnotesize 00010} & {\footnotesize $Z_{3}$} & {\footnotesize 
$|\zeta_{2}\rangle=\frac{1}{2}(|00000\rangle-|00011\rangle+|11100\rangle-|11111\rangle)$} & {\footnotesize 
$|\xi^{+}\rangle_{AB}|\psi^{-}\rangle_{AB}$}\tabularnewline
\hline
{\footnotesize 00011} & {\footnotesize $Z_{2}Z_{3}$} & {\footnotesize 
$|\zeta_{3}\rangle=\frac{1}{2}(|00000\rangle-|00011\rangle-|11100\rangle+|11111\rangle)$} & {\footnotesize 
$|\xi^{-}\rangle_{AB}|\psi^{-}\rangle_{AB}$}\tabularnewline
\hline
{\footnotesize 00100} & {\footnotesize $X_{1}$} & {\footnotesize 
$|\zeta_{4}\rangle=\frac{1}{2}(|01100\rangle+|01111\rangle+|10000\rangle+|10011\rangle)$} & {\footnotesize 
$|\chi^{+}\rangle_{AB}|\psi^{+}\rangle_{AB}$}\tabularnewline
\hline
{\footnotesize 00101} & {\footnotesize $Z_{2}X_{1}$} & {\footnotesize 
$|\zeta_{5}\rangle=\frac{1}{2}(|01100\rangle+|01111\rangle-|10000\rangle-|10011\rangle)$} & {\footnotesize 
$|\chi^{-}\rangle_{AB}|\psi^{+}\rangle_{AB}$}\tabularnewline
\hline
{\footnotesize 00110} & {\footnotesize $Z_{3}X_{1}$} & {\footnotesize 
$|\zeta_{6}\rangle=\frac{1}{2}(|01100\rangle-|01111\rangle+|10000\rangle-|10011\rangle)$} & {\footnotesize 
$|\chi^{+}\rangle_{AB}|\psi^{-}\rangle_{AB}$}\tabularnewline
\hline
{\footnotesize 00111} & {\footnotesize $Z_{2}Z_{3}X_{1}$} & {\footnotesize 
$|\zeta_{7}\rangle=\frac{1}{2}(|01100\rangle-|01111\rangle-|10000\rangle+|10011\rangle)$} & {\footnotesize 
$|\chi^{-}\rangle_{AB}|\psi^{-}\rangle_{AB}$}\tabularnewline
\hline
{\footnotesize 01000} & {\footnotesize $X_{2}$} & {\footnotesize 
$|\zeta_{8}\rangle=\frac{1}{2}(|01000\rangle+|01011\rangle+|10100\rangle+|10111\rangle)$} & {\footnotesize 
$|\vartheta^{+}\rangle_{AB}|\psi^{+}\rangle_{AB}$}\tabularnewline
\hline
{\footnotesize 01001} & {\footnotesize $Z_{2}X_{2}$} & {\footnotesize 
$|\zeta_{9}\rangle=\frac{1}{2}(|01000\rangle+|01011\rangle-|10100\rangle-|10111\rangle)$} & {\footnotesize 
$|\vartheta^{-}\rangle_{AB}|\psi^{+}\rangle_{AB}$}\tabularnewline
\hline
{\footnotesize 01010} & {\footnotesize $Z_{3}X_{2}$} & {\footnotesize 
$|\zeta_{10}\rangle=\frac{1}{2}(|01000\rangle-|01011\rangle+|10100\rangle-|10111\rangle)$} & {\footnotesize 
$|\vartheta^{+}\rangle_{AB}|\psi^{-}\rangle_{AB}$}\tabularnewline
\hline
{\footnotesize 01011} & {\footnotesize $Z_{2}Z_{3}X_{2}$} & {\footnotesize 
$|\zeta_{11}\rangle=\frac{1}{2}(|01000\rangle-|01011\rangle-|10100\rangle+|10111\rangle)$} & {\footnotesize 
$|\vartheta^{-}\rangle_{AB}|\psi^{-}\rangle_{AB}$}\tabularnewline
\hline
{\footnotesize 01100} & {\footnotesize $X_{1}X_{2}$} & {\footnotesize 
$|\zeta_{12}\rangle=\frac{1}{2}(|00100\rangle+|00111\rangle+|11000\rangle+|11011\rangle)$} & {\footnotesize 
$|\theta^{+}\rangle_{AB}|\psi^{+}\rangle_{AB}$}\tabularnewline
\hline
{\footnotesize 01101} & {\footnotesize $Z_{2}X_{1}X_{2}$} & {\footnotesize 
$|\zeta_{13}\rangle=\frac{1}{2}(|00100\rangle+|00111\rangle-|11000\rangle-|11011\rangle)$} & {\footnotesize 
$|\theta^{-}\rangle_{AB}|\psi^{+}\rangle_{AB}$}\tabularnewline
\hline
{\footnotesize 01110} & {\footnotesize $Z_{3}X_{1}X_{2}$} & {\footnotesize 
$|\zeta_{14}\rangle=\frac{1}{2}(|00100\rangle-|00111\rangle+|11000\rangle-|11011\rangle)$} & {\footnotesize 
$|\theta^{+}\rangle_{AB}|\psi^{-}\rangle_{AB}$}\tabularnewline
\hline
{\footnotesize 01111} & {\footnotesize $Z_{2}Z_{3}X_{1}X_{2}$} & {\footnotesize 
$|\zeta_{15}\rangle=\frac{1}{2}(|00100\rangle-|00111\rangle-|11000\rangle+|11011\rangle)$} & {\footnotesize 
$|\theta^{-}\rangle_{AB}|\psi^{-}\rangle_{AB}$}\tabularnewline
\hline
{\footnotesize 10000} & {\footnotesize $X_{3}$} & {\footnotesize 
$|\zeta_{16}\rangle=\frac{1}{2}(|00001\rangle+|00010\rangle+|11101\rangle+|11110\rangle)$} & {\footnotesize 
$|\xi^{+}\rangle_{AB}|\phi^{+}\rangle_{AB}$}\tabularnewline
\hline
{\footnotesize 10001} & {\footnotesize $Z_{2}X_{3}$} & {\footnotesize 
$|\zeta_{17}\rangle=\frac{1}{2}(|00001\rangle+|00010\rangle-|11101\rangle-|11110\rangle)$} & {\footnotesize 
$|\xi^{-}\rangle_{AB}|\phi^{+}\rangle_{AB}$}\tabularnewline
\hline
{\footnotesize 10010} & {\footnotesize $Z_{3}X_{3}$} & {\footnotesize 
$|\zeta_{18}\rangle=\frac{1}{2}(|00001\rangle-|00010\rangle+|11101\rangle-|11110\rangle)$} & {\footnotesize 
$|\xi^{+}\rangle_{AB}|\phi^{-}\rangle_{AB}$}\tabularnewline
\hline
{\footnotesize 10011} & {\footnotesize $Z_{2}Z_{3}X_{3}$} & {\footnotesize 
$|\zeta_{19}\rangle=\frac{1}{2}(|00001\rangle-|00010\rangle-|11101\rangle+|11110\rangle)$} & {\footnotesize 
$|\xi^{-}\rangle_{AB}|\phi^{-}\rangle_{AB}$}\tabularnewline
\hline
{\footnotesize 10100} & {\footnotesize $X_{1}X_{3}$} & {\footnotesize 
$|\zeta_{20}\rangle=\frac{1}{2}(|01101\rangle+|01110\rangle+|10001\rangle+|10010\rangle)$} & {\footnotesize 
$|\chi^{+}\rangle_{AB}|\phi^{+}\rangle_{AB}$}\tabularnewline
\hline
{\footnotesize 10101} & {\footnotesize $Z_{2}X_{1}X_{3}$} & {\footnotesize 
$|\zeta_{21}\rangle=\frac{1}{2}(|01101\rangle+|01110\rangle-|10001\rangle-|10010\rangle)$} & {\footnotesize 
$|\chi^{-}\rangle_{AB}|\phi^{+}\rangle_{AB}$}\tabularnewline
\hline
{\footnotesize 10110} & {\footnotesize $Z_{3}X_{1}X_{3}$} & {\footnotesize 
$|\zeta_{22}\rangle=\frac{1}{2}(|01101\rangle-|01110\rangle+|10001\rangle-|10010\rangle)$} & {\footnotesize 
$|\chi^{+}\rangle_{AB}|\phi^{-}\rangle_{AB}$}\tabularnewline
\hline
{\footnotesize 10111} & {\footnotesize $Z_{2}Z_{3}X_{1}X_{3}$} & {\footnotesize 
$|\zeta_{23}\rangle=\frac{1}{2}(|01101\rangle-|01110\rangle-|10001\rangle+|10010\rangle)$} & {\footnotesize 
$|\chi^{-}\rangle_{AB}|\phi^{-}\rangle_{AB}$}\tabularnewline
\hline
{\footnotesize 11000} & {\footnotesize $X_{2}X_{3}$} & {\footnotesize 
$|\zeta_{24}\rangle=\frac{1}{2}(|01001\rangle+|01010\rangle+|10101\rangle+|10110\rangle)$} & {\footnotesize 
$|\vartheta^{+}\rangle_{AB}|\phi^{+}\rangle_{AB}$}\tabularnewline
\hline
{\footnotesize 11001} & {\footnotesize $Z_{2}X_{2}X_{3}$} & {\footnotesize 
$|\zeta_{25}\rangle=\frac{1}{2}(|01001\rangle+|01010\rangle-|10101\rangle-|10110\rangle)$} & {\footnotesize 
$|\vartheta^{-}\rangle_{AB}|\phi^{+}\rangle_{AB}$}\tabularnewline
\hline
{\footnotesize 11010} & {\footnotesize $Z_{3}X_{2}X_{3}$} & {\footnotesize 
$|\zeta_{26}\rangle=\frac{1}{2}(|01001\rangle-|01010\rangle+|10101\rangle-|10110\rangle)$} & {\footnotesize 
$|\vartheta^{+}\rangle_{AB}|\phi^{-}\rangle_{AB}$}\tabularnewline
\hline
{\footnotesize 11011} & {\footnotesize $Z_{2}Z_{3}X_{2}X_{3}$} & {\footnotesize 
$|\zeta_{27}\rangle=\frac{1}{2}(|01001\rangle-|01010\rangle-|10101\rangle+|10110\rangle)$} & {\footnotesize 
$|\vartheta^{-}\rangle_{AB}|\phi^{-}\rangle_{AB}$}\tabularnewline
\hline
{\footnotesize 11100} & {\footnotesize $X_{1}X_{2}X_{3}$} & {\footnotesize 
$|\zeta_{28}\rangle=\frac{1}{2}(|00101\rangle+|00110\rangle+|11001\rangle+|11010\rangle)$} & {\footnotesize 
$|\theta^{+}\rangle_{AB}|\phi^{+}\rangle_{AB}$}\tabularnewline
\hline
{\footnotesize 11101} & {\footnotesize $Z_{2}X_{1}X_{2}X_{3}$} & {\footnotesize 
$|\zeta_{29}\rangle=\frac{1}{2}(|00101\rangle+|00110\rangle-|11001\rangle-|11010\rangle)$} & {\footnotesize 
$|\theta^{-}\rangle_{AB}|\phi^{+}\rangle_{AB}$}\tabularnewline
\hline
{\footnotesize 11110} & {\footnotesize $Z_{3}X_{1}X_{2}X_{3}$} & {\footnotesize 
$|\zeta_{30}\rangle=\frac{1}{2}(|00101\rangle-|00110\rangle+|11001\rangle-|11010\rangle)$} & {\footnotesize 
$|\theta^{+}\rangle_{AB}|\phi^{-}\rangle_{AB}$}\tabularnewline
\hline
{\footnotesize 11111} & {\footnotesize $Z_{2}Z_{3}X_{1}X_{2}X_{3}$} & {\footnotesize 
$|\zeta_{31}\rangle=\frac{1}{2}(|00101\rangle-|00110\rangle-|11001\rangle+|11010\rangle)$} & {\footnotesize 
$|\theta^{-}\rangle_{AB}|\phi^{-}\rangle_{AB}$}\tabularnewline
\hline
\end{tabular}

~

\subsection*{~~~~~~~~~~~~~~~~~~~~~~~~~~~~~~~~~~~~~~~~{\normalsize III.
TELEPORTATION }}

We now proceed to the implementation of teleportation protocols, starting
with the teleportation of single and two qubit states. Subsequently,
we illustrate the protocol of arbitrary $N$-qubit teleportation.

\subsubsection*{Teleportation of single qubit state}

For the purpose of single qubit teleportation, Alice and Bob are assigned
to be in the state: 
$|\zeta\rangle=\frac{1}{\sqrt{2}}(|0_{A}0_{A}0_{A}\rangle+|1_{A}1_{A}1_{A}\rangle)\frac{1}{\sqrt{2}}(|0_{A}0_{B}\rangle+|1_{A}1_{B}\rangle)$.
For teleporting an arbitrary single qubit state $(\alpha|0\rangle+\beta|1\rangle)$,
where $|\alpha|^{2}+|\beta|^{2}=1$, the combined six-qubit state
obtained by Alice can be written as,

$(\alpha|0\rangle+\beta|1\rangle)\frac{1}{2}(|00000\rangle+|00011\rangle+|11100\rangle+|11111\rangle)$

$=\frac{1}{2}(|00000\rangle+|01110\rangle+|10001\rangle+|11111\rangle)(\alpha|0\rangle+\beta|1\rangle)+\frac{1}{2}(|00000\rangle+|01110\rangle-|10001\rangle-|11111\rangle)(\alpha|0\rangle-\beta|1\rangle)+$

$\frac{1}{2}(|00001\rangle+|01111\rangle+|10000\rangle+|11110\rangle)(\alpha|1\rangle+\beta|0\rangle)+\frac{1}{2}(|00001\rangle+|01111\rangle-|10000\rangle-|11110\rangle)(\alpha|1\rangle-\beta|0\rangle)$.~~~~~~~~~~~~(5)
Making a von Neumann five-particle measurement using the orthonormal
states given in the first column of Table II, Alice sends her outcome
to Bob via two classical bits. Subsequently, Bob applies suitable
unitary operations to recover the original state $(\alpha|0\rangle+\beta|1\rangle)$.
Hence, the teleportation protocol is deterministically implemented.

\subsubsection*{{\footnotesize Table II: The outcome of the measurement performed
by Alice and the stats obtained by Bob.}}

~~~~~~~~~~~~~~~~~~~~~~~~~~~~~~~\begin{tabular}{|c|c|}
\hline
{\footnotesize Outcome of the measurement} & {\footnotesize State obtained}\tabularnewline
\hline
\hline
{\footnotesize $\frac{1}{2}(|00000\rangle+|01110\rangle+|10001\rangle+|11111\rangle)$} & {\footnotesize 
$\alpha|0\rangle+\beta|1\rangle$}\tabularnewline
\hline
{\footnotesize $\frac{1}{2}(|00000\rangle+|01110\rangle-|10001\rangle-|11111\rangle)$} & {\footnotesize 
$\alpha|0\rangle-\beta|1\rangle$}\tabularnewline
\hline
{\footnotesize $\frac{1}{2}(|00001\rangle+|01111\rangle+|10000\rangle+|11110\rangle)$} & {\footnotesize 
$\alpha|1\rangle+\beta|0\rangle$}\tabularnewline
\hline
{\footnotesize $\frac{1}{2}(|00001\rangle+|01111\rangle-|10000\rangle-|11110\rangle)$} & {\footnotesize 
$\alpha|1\rangle-\beta|0\rangle$}\tabularnewline
\hline
\end{tabular}

~~

\subsubsection*{Teleportation of arbritrary two-qubit state}

For two qubit teleportation, Alice and Bob share the state 
$\frac{1}{2}(|0_{A}0_{A}0_{B}\rangle+|1_{A}1_{A}1_{B}\rangle)(|0_{A}0_{B}\rangle+|1_{A}1_{B}\rangle)$,
where Alice possesses three qubits and Bob possesses two. The combined
seven-qubit state containing an arbitrary state $(\alpha|00\rangle+\gamma|01\rangle+\mu|10\rangle+\beta|11\rangle)$
with $|\alpha|^{2}+|\beta|^{2}+|\gamma|^{2}+|\mu|^{2}=1$, can be
written as the linear combination of sixteen states which are decomposed
into Alice and Bob system:

$(\alpha|00\rangle+\gamma|01\rangle+\mu|10\rangle+\beta|11\rangle)\frac{1}{2}(|00000\rangle+|00011\rangle+|11100\rangle+|11111\rangle)$

$=|\Omega_{0}\rangle(\alpha|00\rangle+\gamma|01\rangle+\mu|10\rangle+\beta|11\rangle)+|\Omega_{1}\rangle(\alpha|00\rangle+\gamma|01\rangle-\mu|10\rangle-\beta|11\rangle)+$$|\Omega_{2}\rangle(\alpha|00\rangle-\gamma|01\rangle+\mu|10\rangle-\beta|11\rangle)$

$+|\Omega_{3}\rangle(\alpha|00\rangle-\gamma|01\rangle-\mu|10\rangle+\beta|11\rangle)+$$|\Omega_{4}\rangle(\alpha|10\rangle+\gamma|11\rangle+\mu|00\rangle+\beta|01\rangle)+|\Omega_{5}\rangle(\alpha|10\rangle+\gamma|11\rangle-\mu|00\rangle-\beta|01\rangle)$

$+|\Omega_{6}\rangle(\alpha|10\rangle-\gamma|11\rangle+\mu|00\rangle-\beta|01\rangle)+|\Omega_{7}\rangle(\alpha|10\rangle-\gamma|11\rangle-\mu|00\rangle+\beta|01\rangle)+$$|\Omega_{8}\rangle(\alpha|01\rangle+\gamma|00\rangle+\mu|11\rangle+\beta|10\rangle)$

$+|\Omega_{9}\rangle(\alpha|01\rangle+\gamma|00\rangle-\mu|11\rangle-\beta|10\rangle)+$$|\Omega_{10}\rangle(\alpha|01\rangle-\gamma|00\rangle+\mu|11\rangle-\beta|10\rangle)+|\Omega_{11}\rangle(\alpha|01\rangle-\gamma|00\rangle-\mu|11\rangle+\beta|10\rangle)$

$+|\Omega_{12}\rangle(\alpha|11\rangle+\gamma|10\rangle+\mu|01\rangle+\beta|00\rangle)+|\Omega_{13}\rangle(\alpha|11\rangle+\gamma|10\rangle-\mu|01\rangle-\beta|00\rangle)+|\Omega_{14}\rangle(\alpha|11\rangle-\gamma|10\rangle+\mu|01\rangle-\beta|00\rangle)$

$+|\Omega_{15}\rangle(\alpha|11\rangle-\gamma|10\rangle-\mu|01\rangle+\beta|00\rangle)$.~~~~~~~~~~~~~~~~~~~~~~~~~~~~~~~~~~~~~~~~~~~~~~~~~~~~~~~~~~~~~~~~~~~~~~~~~~~~~~~~~~~~~~~~~~~~~~~~~~~~~~~~~~~~~~~~~~~~~~~~~~~~~~~~~~~~~~~~~~(6)
where $|\Omega_{i}\rangle's$ ($i=0,1,2,...,15$) are mutually orthonormal
five particle states, belonging to Alice, as shown in Table III. Now
Alice makes a five-particle measurement using those orthonormal states
and sends the outcome of her measurement to Bob via four classical
bits. After applying suitable unitary operations Bob will be able
to reconstruct the original state $(\alpha|00\rangle+\gamma|01\rangle+\mu|10\rangle+\beta|11\rangle)$.
All the possible states obtained after Alice's measurement are orthogonal,
so it completes the deterministic teleportation protocol.

\subsubsection*{{\footnotesize Table III: The outcome of the measurement i.e.,}{\small{}
$|\Omega_{i}\rangle's$ ($i=0,1,2,...,15$)}{\footnotesize{} performed
by Alice and the state obtained by Bob, where 1,2,3,4,5 in $|\Omega_{i}\rangle_{1,2,3,4,5}$
denote the order of the Alice's five particle state from left hand
side.}}

~~~~~~~~~~~~~~~~~~~~~~~~~~~~~~\begin{tabular}{|c|c|}
\hline
{\footnotesize Outcome of the measurement} & {\footnotesize State obtained}\tabularnewline
\hline
\hline
{\footnotesize $|\Omega_{0}\rangle_{1,2,3,4,5}=\frac{1}{2}(|00000\rangle+|01001\rangle+|10110\rangle+|11111\rangle)$} & 
{\footnotesize $\alpha|00\rangle+\gamma|01\rangle+\mu|10\rangle+\beta|11\rangle$}\tabularnewline
\hline
{\footnotesize $|\Omega_{1}\rangle_{1,2,3,4,5}=\frac{1}{2}(|00000\rangle+|01001\rangle-|10110\rangle-|11111\rangle)$} & 
{\footnotesize $\alpha|00\rangle+\gamma|01\rangle-\mu|10\rangle-\beta|11\rangle$}\tabularnewline
\hline
{\footnotesize $|\Omega_{2}\rangle_{1,2,3,4,5}=\frac{1}{2}(|00000\rangle-|01001\rangle+|10110\rangle-|11111\rangle)$} & 
{\footnotesize $\alpha|00\rangle-\gamma|01\rangle+\mu|10\rangle-\beta|11\rangle$}\tabularnewline
\hline
{\footnotesize $|\Omega_{3}\rangle_{1,2,3,4,5}=\frac{1}{2}(|00000\rangle-|01001\rangle-|10110\rangle+|11111\rangle)$} & 
{\footnotesize $\alpha|00\rangle-\gamma|01\rangle-\mu|10\rangle+\beta|11\rangle$}\tabularnewline
\hline
{\footnotesize $|\Omega_{4}\rangle_{1,2,3,4,5}=\frac{1}{2}(|00110\rangle+|01111\rangle+|10000\rangle+|11001\rangle)$} & 
{\footnotesize $\alpha|10\rangle+\gamma|11\rangle+\mu|00\rangle+\beta|01\rangle$}\tabularnewline
\hline
{\footnotesize $|\Omega_{5}\rangle_{1,2,3,4,5}=\frac{1}{2}(|00110\rangle+|01111\rangle-|10000\rangle-|11001\rangle)$} & 
{\footnotesize $\alpha|10\rangle+\gamma|11\rangle-\mu|00\rangle-\beta|01\rangle$}\tabularnewline
\hline
{\footnotesize $|\Omega_{6}\rangle_{1,2,3,4,5}=\frac{1}{2}(|00110\rangle-|01111\rangle+|10000\rangle-|11001\rangle)$} & 
{\footnotesize $\alpha|10\rangle-\gamma|11\rangle+\mu|00\rangle-\beta|01\rangle$}\tabularnewline
\hline
{\footnotesize $|\Omega_{7}\rangle_{1,2,3,4,5}=\frac{1}{2}(|00110\rangle-|01111\rangle-|10000\rangle+|11001\rangle)$} & 
{\footnotesize $\alpha|10\rangle-\gamma|11\rangle-\mu|00\rangle+\beta|01\rangle$}\tabularnewline
\hline
{\footnotesize $|\Omega_{8}\rangle_{1,2,3,4,5}=\frac{1}{2}(|00001\rangle+|01000\rangle+|10111\rangle+|11110\rangle)$} & 
{\footnotesize $\alpha|01\rangle+\gamma|00\rangle+\mu|11\rangle+\beta|10\rangle$}\tabularnewline
\hline
{\footnotesize $|\Omega_{9}\rangle_{1,2,3,4,5}=\frac{1}{2}(|00001\rangle+|01000\rangle-|10111\rangle-|11110\rangle)$} & 
{\footnotesize $\alpha|01\rangle+\gamma|00\rangle-\mu|11\rangle-\beta|10\rangle$}\tabularnewline
\hline
{\footnotesize $|\Omega_{10}\rangle_{1,2,3,4,5}=\frac{1}{2}(|00001\rangle-|01000\rangle+|10111\rangle-|11110\rangle)$} & 
{\footnotesize $\alpha|01\rangle-\gamma|00\rangle+\mu|11\rangle-\beta|10\rangle$}\tabularnewline
\hline
{\footnotesize $|\Omega_{11}\rangle_{1,2,3,4,5}=\frac{1}{2}(|00001\rangle-|01000\rangle-|10111\rangle+|11110\rangle)$} & 
{\footnotesize $\alpha|01\rangle-\gamma|00\rangle-\mu|11\rangle+\beta|10\rangle$}\tabularnewline
\hline
{\footnotesize $|\Omega_{12}\rangle_{1,2,3,4,5}=\frac{1}{2}(|00111\rangle+|01110\rangle+|10001\rangle+|11000\rangle)$} & 
{\footnotesize $\alpha|11\rangle+\gamma|10\rangle+\mu|01\rangle+\beta|00\rangle$}\tabularnewline
\hline
{\footnotesize $|\Omega_{13}\rangle_{1,2,3,4,5}=\frac{1}{2}(|00111\rangle+|01110\rangle-|10001\rangle-|11000\rangle)$} & 
{\footnotesize $\alpha|11\rangle+\gamma|10\rangle-\mu|01\rangle-\beta|00\rangle$}\tabularnewline
\hline
{\footnotesize $|\Omega_{14}\rangle_{1,2,3,4,5}=\frac{1}{2}(|00111\rangle-|01110\rangle+|10001\rangle-|11000\rangle)$} & 
{\footnotesize $\alpha|11\rangle-\gamma|10\rangle+\mu|01\rangle-\beta|00\rangle$}\tabularnewline
\hline
{\footnotesize $|\Omega_{15}\rangle_{1,2,3,4,5}=\frac{1}{2}(|00111\rangle-|01110\rangle-|10001\rangle+|11000\rangle)$} & 
{\footnotesize $\alpha|11\rangle-\gamma|10\rangle-\mu|01\rangle+\beta|00\rangle$}\tabularnewline
\hline
\end{tabular}

~~

By rearranging the order of the particles we can write the same state
as, 
$|\Omega_{0}\rangle_{1,4,3,2,5}=\frac{1}{2}(|00000\rangle+|00011\rangle+|11100\rangle+|11111\rangle)=|\xi^{+}\rangle|\psi^{+}\rangle$.
Simillarly, $|\Omega_{1}\rangle_{1,4,3,2,5}=|\zeta^{-}\rangle|\psi^{+}\rangle$,
$|\Omega_{2}\rangle_{1,4,3,2,5}=|\zeta^{+}\rangle|\psi^{-}\rangle$,
$|\Omega_{3}\rangle_{1,4,3,2,5}=|\zeta^{-}\rangle|\psi^{-}\rangle$,
$|\Omega_{4}\rangle_{1,4,3,2,5}=|\chi^{+}\rangle|\psi^{+}\rangle$
etc.

\subsubsection*{Generalized teleportation of arbitrary $N$-qubit state}

To achieve the purpose of $N$-qubit state teleportation, we have
to start with the $(2N+1)$-qubit state $|\zeta_{0}\rangle$ given
in the Eq. 3, such that Alice possesses `$N+1$' qubits and Bob possesses
`$N$' qubits. The arbitrary $N$-qubit state which Alice wants to
teleport to Bob is, $\Sigma_{i=0}^{2^{N}-1}\alpha_{i}|a_{i}\rangle$
satisfying $\Sigma_{i=0}^{2^{N}-1}|\alpha_{i}|^{2}=1.$ Here $a_{i}$
is binary representation of `$i$'. The combined state can be written
as the linear combination of $2^{2N}$ number of states which are
decomposed into Alice and Bob system:

$(\Sigma_{i=0}^{2^{N}-1}\alpha_{i}|a_{i}\rangle)|\zeta_{0}\rangle=\Sigma_{j=0}^{2^{2N}-1}|\Omega_{j}\rangle_{A}|\eta_{j}\rangle_{B}$,
such that $|\Omega_{j}\rangle_{A}'s$ ($j=0,1,2,....,2^{2N}-1$) are
mutually orthogonal states. We now explicate how to obtain $|\Omega_{j}\rangle's$
for any value of $N$. For any given $N$, $|\Omega_{0}\rangle$ is
the product of one $|\xi^{+}\rangle$ and ($N-1$) number of $|\psi^{+}\rangle$
states with suitable rearrangement of the particles:

For even $N$, $|\Omega_{0}\rangle_{1,4,3,2,7,...,2N,2N-3,2N-2,2N+1}=|\zeta^{+}\rangle|\psi^{+}\rangle........|\psi^{+}\rangle$
and

for odd $N$, $|\Omega_{0}\rangle_{1,2,5,4,3,...,2N,2N-3,2N-2,2N+1}=|\zeta^{+}\rangle|\psi^{+}\rangle........|\psi^{+}\rangle$
, where $1,2,3,...,2N,2N+1$ are the order of the particles in the
state, 
$|\Omega_{0}\rangle_{A}$.~~~~~~~~~~~~~~~~~~~~~~~~~~~~~~~~~~~~~~~~~~~~~~~~~~~~~~~~~~~~~~~~~~~~~~~~~~~~~~~~~~~~~~~~~~~~~~~~~~~~~~~~~~~~~~~~~~~~~~~~~~~~~~~~~~~~~~~~~~~~~(7)
Once, the $|\Omega_{0}\rangle$ is known we can calculate other $|\Omega_{j}\rangle$'s
:

$|\Omega_{j}\rangle_{1,2,3...,2N,2N+1}=$$\otimes_{k=1}^{N}(Z_{k})^{b_{k}}(X_{k})^{b_{k+N}}|\Omega_{0}\rangle_{1,2,3...,2N,2N+1}$~~~~~~~~~~~~~~~~~~~~~~~~~~~~~~~~~~~~~~~~~~~~~~~~~~~~~~~~~
~~~~~~~~~~~~~~~~~~~~~~~~~~~~~~~(8)
where $j$ is the decimal representation of a binary bit string $b_{2N}.......b_{2}b_{1}$
($b_{k}=0$ or $1$). After the measurement using these orthogonal
states, Alice's state evolves into one of the $|\Omega_{j}\rangle's$,
which ensures that this protocol is deterministic. If Alice's state
collapses to $|\Omega_{j}\rangle$, then Bob has to apply $\otimes_{k=1}^{N}(Z_{k})^{b_{k}}(X_{k})^{b_{k+N}}$
on his $N$-qubit system to get the unknown state, where $j$ is the
decimal representation of a binary bit string $b_{2N}.......b_{2}b_{1}$.

For example, if we consider the two-qubit state teleportation, then
from Eq. 7 we get, $|\Omega_{0}\rangle_{1,4,3,2,5}=|\zeta^{+}\rangle|\psi^{+}\rangle$
or, $|\Omega_{0}\rangle_{1,2,3,4,5}=\frac{1}{2}(|00000\rangle+|01001\rangle+|10110\rangle+|11111\rangle)$.
Suppose, we want to obtain $|\Omega_{12}\rangle$. We know that 1100
is the binary representation of 12, so using Eq. 8, one can get, 
$|\Omega_{12}\rangle=(X_{1})^{1}(Z_{1})^{0}(X_{2})^{1}(Z_{2})^{0}|\Omega_{0}\rangle=(X_{1})(X_{2})|\Omega_{0}\rangle=\frac{1}{2}(|11000\rangle+|10001\rangle+|01110\rangle+|00111\rangle)$,
which exactly matches with the $|\Omega_{12}\rangle$ in Table III.
So after the measurement, if Alice's state collapses to $|\Omega_{12}\rangle$,
then Bob has to apply $(X_{1})(X_{2})$ on his state to reconstruct
the unknown state.

~~

\subsection*{~~~~~~~~~~~~~~~~~~~~~~~~~~~~{\normalsize IV.
QUANTUM INFORMATION SPLITTING }}

It can be shown that, the five-qubit state $|\zeta\rangle$ can be
used for single qubit QIS through three different protocols, where
as only one protocol is feasible for two-qubit QIS {[}36{]}.

\subsubsection*{QIS of a single qubit state}

The distributions of the qubits, within the three parties of the initial
state $|\zeta\rangle$ are different for three different protocols
of single qubit QIS. Those states are:

(i) 
$\frac{1}{\sqrt{2}}(|0_{A}0_{B}0_{B}\rangle+|1_{A}1_{B}1_{B}\rangle)\frac{1}{\sqrt{2}}(|0_{B}0_{C}\rangle+|1_{B}1_{C}\rangle)$,

(ii) 
$\frac{1}{\sqrt{2}}(|0_{A}0_{A}0_{B}\rangle+|1_{A}1_{A}1_{B}\rangle)\frac{1}{\sqrt{2}}(|0_{B}0_{C}\rangle+|1_{B}1_{C}\rangle)$,

(iii) 
$\frac{1}{\sqrt{2}}(|0_{A}0_{A}0_{B}\rangle+|1_{A}1_{A}1_{B}\rangle)\frac{1}{\sqrt{2}}(|0_{A}0_{C}\rangle+|1_{A}1_{C}\rangle)$.

It is noted that Alice possesses one, two and three qubits in three
different initial states and Charlie possesses one qubit in all those
states. We first explicitly demostrate this protocol using the state
$\frac{1}{\sqrt{2}}(|0_{A}0_{B}0_{B}\rangle+|1_{A}1_{B}1_{B}\rangle)\frac{1}{\sqrt{2}}(|0_{B}0_{C}\rangle+|1_{B}1_{C}\rangle)$
and then we discuss about the other two protocols. The combined state
of $\frac{1}{\sqrt{2}}(|0_{A}0_{B}0_{B}\rangle+|1_{A}1_{B}1_{B}\rangle)\frac{1}{\sqrt{2}}(|0_{B}0_{C}\rangle+|1_{B}1_{C}\rangle)$
with the unknown qubit ($\alpha|0\rangle+\beta|1\rangle$) where $|\alpha|^{2}+|\beta|^{2}=1$,
can be written as,

$(\alpha|0\rangle+\beta|1\rangle)\frac{1}{2}(|00000\rangle+|00011\rangle+|11100\rangle+|11111\rangle)$

$=\frac{1}{2}\{|00\rangle\alpha(|0000\rangle+|0011\rangle)+|01\rangle\alpha(|1100\rangle+|1111\rangle)+|10\rangle\beta(|0000\rangle+|0011\rangle)+|11\rangle\beta(|1100\rangle+|1111\rangle)\}$.~~~~~~~~~~~~~~~~~~~~~~(9)
Alice first performs a measurement using Bell pairs such that Bob
and Charlie evolve into an entangled state given in Table IV and conveys
her outcome to Charlie by two cbits.

\subsubsection*{{\footnotesize Table IV: The outcome of the measurement performed
by Alice and the entangled state obtained by Bob and Charlie (the
coefficients are removed for convenience).}}

{\footnotesize ~~~~~~~~~~~~~~~~~~~~~~~~~~~~~~~~~~~~~}\begin{tabular}{|c|c|}
\hline
{\footnotesize Outcome of the measurement} & {\footnotesize Entangled state obtained by Bob and Charlie}\tabularnewline
\hline
\hline
{\footnotesize $|00\rangle+|11\rangle$} & {\footnotesize 
$\alpha(|0000\rangle+|0011\rangle)+\beta(|1100\rangle+|1111\rangle)$}\tabularnewline
\hline
{\footnotesize $|01\rangle+|10\rangle$} & {\footnotesize 
$\alpha(|1100\rangle+|1111\rangle)+\beta(|0000\rangle+|0011\rangle)$}\tabularnewline
\hline
{\footnotesize $|01\rangle-|10\rangle$} & {\footnotesize 
$\alpha(|1100\rangle+|1111\rangle)-\beta(|0000\rangle+|0011\rangle)$}\tabularnewline
\hline
{\footnotesize $|00\rangle-|11\rangle$} & {\footnotesize 
$\alpha(|0000\rangle+|0011\rangle)-\beta(|1100\rangle+|1111\rangle)$}\tabularnewline
\hline
\end{tabular}

~

Bob then performs a three-particle measurement and conveys his outcome
to Charlie via two cbits of information. Having known the outcomes
of both their measurements, Charlie can obtain the unknown qubit,
by performing a suitable unitary operation on his qubit. Hence, the
QIS protocol is satisfied. Suppose, after the measurement of Alice,
the Bob-Charlie system evolves into the state $\alpha(|0000\rangle+|0011\rangle)+\beta(|1100\rangle+|1111\rangle)$,
then the outcome of the measurement performed by Bob and the state
obtained by Charlie are shown in Table V.

\subsubsection*{{\footnotesize Table V: The possible outcomes of the measurement
performed by Bob and the state obtained by Charlie.}}

{\footnotesize ~~~~~~~~~~~~~~~~~~~~~~~~~~~~~~~~~~~~~~~~}\begin{tabular}{|c|c|}
\hline
{\footnotesize Outcome of the measurement} & {\footnotesize State obtained}\tabularnewline
\hline
\hline
{\footnotesize $|000\rangle+|111\rangle$} & {\footnotesize $\alpha|0\rangle+\beta|1\rangle$}\tabularnewline
\hline
{\footnotesize $|001\rangle+|110\rangle$} & {\footnotesize $\alpha|1\rangle+\beta|0\rangle$}\tabularnewline
\hline
{\footnotesize $|000\rangle-|111\rangle$} & {\footnotesize $\alpha|0\rangle-\beta|1\rangle$}\tabularnewline
\hline
{\footnotesize $|001\rangle-|110\rangle$} & {\footnotesize $\alpha|1\rangle-\beta|0\rangle$}\tabularnewline
\hline
\end{tabular}

For the another protocol Alice, Bob and Charlie share a state, 
$\frac{1}{\sqrt{2}}(|0_{A}0_{A}0_{B}\rangle+|1_{A}1_{A}1_{B}\rangle)\frac{1}{\sqrt{2}}(|0_{B}0_{C}\rangle+|1_{B}1_{C}\rangle)$,
so the combined state containing the unknown qubit will be,

$(\alpha|0\rangle+\beta|1\rangle)\frac{1}{2}(|00000\rangle+|00011\rangle+|11100\rangle+|11111\rangle)$

$=|000\rangle\alpha(|000\rangle+|011\rangle)+|011\rangle\alpha(|100\rangle+|111\rangle)+|100\rangle\beta(|000\rangle+|011\rangle)+|111\rangle\beta(|100\rangle+|111\rangle)$.~~~~~~~~~~~~~
~~~~~~~~~~~~~~~(10) Alice first performs measurement
using \{$|000\rangle\pm|111\rangle,|011\rangle\pm|100\rangle,|001\rangle\pm|110\rangle,|010\rangle\pm|101\rangle$\}
and subsequently Bob performs a Bell measurement. Then both of them
convey their outcomes to Charlie. It can be shown that after Bob's
measurement, Charlie's state collapses into $(\alpha|0\rangle\pm\beta|1\rangle)$
or $(\alpha|1\rangle\pm\beta|0\rangle)$. Knowing the outcomes of
those measurements classically, Charlie can reconstruct the information.

Simillarly, in the third protocol, where Alice, Bob and Charlie are
sharing the state $\frac{1}{\sqrt{2}}(|0_{A}0_{A}0_{B}\rangle+|1_{A}1_{A}1_{B}\rangle)$
$\frac{1}{\sqrt{2}}(|0_{A}0_{C}\rangle+|1_{A}1_{C}\rangle)$, Alice
has to perform four particle measurement using the orthogonal states
$(|0000\rangle\pm|1111\rangle),$ $(|0110\rangle\pm|1001\rangle),$$(|0001\rangle\pm|1110\rangle),(|1000\rangle\pm|0111\rangle)$
and then Bob has to perform one particle measurement using 
\{$\frac{|0\rangle+|1\rangle}{\sqrt{2}},\frac{|0\rangle-|1\rangle}{\sqrt{2}}$\}.
After that, Charlie will be able to get the unknown qubit by applying
unitary operations according to the outcomes of those measurements.

\subsubsection*{QIS of two-qubit state}

We now procced to show QIS of two-qubit state where Alice, Bob and
Charlie are sharing the state:

$|\zeta\rangle=\frac{1}{\sqrt{2}}(|0_{A}0_{B}0_{C}\rangle+|1_{A}1_{B}1_{C}\rangle)\frac{1}{\sqrt{2}}(|0_{A}0_{C}\rangle+|1_{A}1_{C}\rangle)$.

Alice combines the unknown qubit $(\alpha|00\rangle+\gamma|01\rangle+\mu|10\rangle+\beta|11\rangle)$,
($|\alpha|^{2}+|\beta|^{2}+|\gamma|^{2}+|\mu|^{2}=1$) with the state
and performs a four-particle measurement using the orthonormal states
given in the first column of Table VI. All the orthonormal states
which form the basis of Alice's measurement can be broken down into
two Bell pairs using suitable rearrangement of particles. Depending
on the outcome of the measurement, Bob and Charlie evolve to an entangled
state. After that, Bob performs a single particle measurement using
Hadamard basis and sends his outcome to Charlie.

Having known the outcomes of both their measurements, Charlie can
obtain the unknown two qubit state, by performing a suitable unitary
operation on his qubit. For instance, after the measurement of Alice,
the Bob-Charlie system collapses to the state $(\alpha|000\rangle+\gamma|001\rangle+\mu|110\rangle+\beta|111\rangle)$.
Then the possible outcomes of Bob's measurement are $\frac{1}{\sqrt{2}}(|0\rangle+|1\rangle)$
or $\frac{1}{\sqrt{2}}(|0\rangle-|1\rangle)$ and the states obtained
by Charlie are $(\alpha|00\rangle+\gamma|01\rangle+\mu|10\rangle+\beta|11\rangle)$
or $(\alpha|00\rangle+\gamma|01\rangle-\mu|10\rangle-\beta|11\rangle)$
respectively, from which the original state can be recovered.

\subsubsection*{{\footnotesize Table VI: The outcome of the measurement performed
by Alice and the entangled state obtained by Bob and Charlie.}}

{\footnotesize ~~~~~~~~~~~~~~~~~~~~~~~~~~~~~~~~}\begin{tabular}{|c|c|}
\hline
{\footnotesize Outcome of the measurement} & {\footnotesize Entangled state obtained by Bob and Charlie}\tabularnewline
\hline
\hline
{\footnotesize $\frac{1}{2}(|0000\rangle+|0101\rangle+|1010\rangle+|1111\rangle)$} & {\footnotesize 
$\alpha|000\rangle+\gamma|001\rangle+\mu|110\rangle+\beta|111\rangle$}\tabularnewline
\hline
{\footnotesize $\frac{1}{2}(|0000\rangle-|0101\rangle+|1010\rangle-|1111\rangle)$} & {\footnotesize 
$\alpha|000\rangle-\gamma|001\rangle+\mu|110\rangle-\beta|111\rangle$}\tabularnewline
\hline
{\footnotesize $\frac{1}{2}(|0000\rangle+|0101\rangle-|1010\rangle-|1111\rangle)$} & {\footnotesize 
$\alpha|000\rangle+\gamma|001\rangle-\mu|110\rangle-\beta|111\rangle$}\tabularnewline
\hline
{\footnotesize $\frac{1}{2}(|0000\rangle-|0101\rangle-|1010\rangle+|1111\rangle)$} & {\footnotesize 
$\alpha|000\rangle-\gamma|001\rangle-\mu|110\rangle+\beta|111\rangle$}\tabularnewline
\hline
{\footnotesize $\frac{1}{2}(|0001\rangle+|0100\rangle+|1011\rangle+|1110\rangle)$} & {\footnotesize 
$\alpha|001\rangle+\gamma|000\rangle+\mu|110\rangle+\beta|110\rangle$}\tabularnewline
\hline
{\footnotesize $\frac{1}{2}(|0001\rangle-|0100\rangle+|1011\rangle-|1110\rangle)$} & {\footnotesize 
$\alpha|001\rangle-\gamma|000\rangle+\mu|110\rangle-\beta|110\rangle$}\tabularnewline
\hline
{\footnotesize $\frac{1}{2}(|0001\rangle+|0100\rangle-|1011\rangle-|1110\rangle)$} & {\footnotesize 
$\alpha|001\rangle+\gamma|000\rangle-\mu|110\rangle-\beta|110\rangle$}\tabularnewline
\hline
{\footnotesize $\frac{1}{2}(|0001\rangle-|0100\rangle-|1011\rangle+|1110\rangle)$} & {\footnotesize 
$\alpha|001\rangle-\gamma|000\rangle-\mu|110\rangle+\beta|110\rangle$}\tabularnewline
\hline
{\footnotesize $\frac{1}{2}(|0010\rangle+|0111\rangle+|1000\rangle+|1101\rangle)$} & {\footnotesize 
$\alpha|110\rangle+\gamma|111\rangle+\mu|000\rangle+\beta|001\rangle$}\tabularnewline
\hline
{\footnotesize $\frac{1}{2}(|0010\rangle-|0111\rangle+|1000\rangle-|1101\rangle)$} & {\footnotesize 
$\alpha|110\rangle-\gamma|111\rangle+\mu|000\rangle-\beta|001\rangle$}\tabularnewline
\hline
{\footnotesize $\frac{1}{2}(|0010\rangle+|0111\rangle-|1000\rangle-|1101\rangle)$} & {\footnotesize 
$\alpha|110\rangle+\gamma|111\rangle-\mu|000\rangle-\beta|001\rangle$}\tabularnewline
\hline
{\footnotesize $\frac{1}{2}(|0010\rangle-|0111\rangle-|1000\rangle+|1101\rangle)$} & {\footnotesize 
$\alpha|110\rangle-\gamma|111\rangle-\mu|000\rangle+\beta|001\rangle$}\tabularnewline
\hline
{\footnotesize $\frac{1}{2}(|0011\rangle+|0110\rangle+|1001\rangle+|1100\rangle)$} & {\footnotesize 
$\alpha|111\rangle+\gamma|110\rangle+\mu|001\rangle+\beta|000\rangle$}\tabularnewline
\hline
{\footnotesize $\frac{1}{2}(|0011\rangle-|0110\rangle+|1001\rangle-|1100\rangle)$} & {\footnotesize 
$\alpha|111\rangle-\gamma|110\rangle+\mu|001\rangle-\beta|000\rangle$}\tabularnewline
\hline
{\footnotesize $\frac{1}{2}(|0011\rangle+|0110\rangle-|1001\rangle-|1100\rangle)$} & {\footnotesize 
$\alpha|111\rangle+\gamma|110\rangle-\mu|001\rangle-\beta|000\rangle$}\tabularnewline
\hline
{\footnotesize $\frac{1}{2}(|0011\rangle-|0110\rangle-|1001\rangle+|1100\rangle)$} & {\footnotesize 
$\alpha|111\rangle-\gamma|110\rangle-\mu|001\rangle+\beta|000\rangle$}\tabularnewline
\hline
\end{tabular}

~~

\subsubsection*{Generalized QIS of arbitrary $N$-qubit state }

For `$N$' qubit QIS Alice, Bob and Charlie need to share a multipartite
entangled state, which is the product of one $|\xi^{+}\rangle$ and
($N-1$) number of $|\psi^{+}\rangle$ states. The state is given
by, $|\zeta^{'}\rangle=|\xi^{+}\rangle_{ABC}|\psi^{+}\rangle_{AC}........|\psi^{+}\rangle_{AC}$
, where $|\xi^{+}\rangle_{ABC}=\frac{1}{\sqrt{2}}(|0_{A}0_{B}0_{C}\rangle+|1_{A}1_{B}1_{C}\rangle)$
and $|\psi^{+}\rangle_{AC}=\frac{1}{\sqrt{2}}(|0_{A}0_{C}\rangle+|1_{A}1_{C}\rangle)$.
The combined system, contaning the unknown state $\Sigma_{i=0}^{2^{N}-1}\alpha_{i}|a_{i}\rangle$
(where $a_{i}$ is `$N$' classical bit string which represents the
binary form of $`i$') can be written as the linear combination of
$2^{2N}$ number of states which are decomposed into Alice's system
and Bob-Charlie combined system:

$(\Sigma_{i=0}^{2^{N}-1}\alpha_{i}|a_{i}\rangle)|\zeta^{'}\rangle=\Sigma_{j=0}^{2^{2N}-1}|\Omega_{j}\rangle_{A}|\eta_{j}\rangle_{BC}$,
where $|\Omega_{j}\rangle_{A}'s$ ($j=0,1,2,....,2^{2N}-1$) are mutually
orthogonal states. We now explicate how to obtain $|\Omega_{j}\rangle's$.
For any given $N$, $|\Omega_{0}\rangle$ is the products of $N$
number of $|\psi^{+}\rangle$ states with suitable rearrangement of
the particles: 
$|\Omega_{0}\rangle_{1,N+1,2,N+2,....,N,2N}=|\psi^{+}\rangle|\psi^{+}\rangle.......|\psi^{+}\rangle$~~~~~~~~~~~~~~~~~~~~~~~~~~~~~~~~~~~~~~~~~~~~~~~~~~~~~~~~~~(11)
where $1,2,3,...,2N,2N+1$ are the order of the particles in the state,
$|\Omega_{0}\rangle_{A}$. Using $|\Omega_{0}\rangle$ we can evaluate
other $|\Omega_{j}\rangle$'s from the equation:

$|\Omega_{j}\rangle_{1,2,3...,2N}=$$\otimes_{k=1}^{N}(Z_{k})^{b_{k}}(X_{k})^{b_{k+N}}|\Omega_{0}\rangle_{1,2,3...,2N}$~~~~~~~~~~~~~~~~~~~~~~~~~~~~~~~~~~~~~~~~~~~~~~~~~~~~~~~~~
~~~~~~~~~~~~~~~~~ ~~~~~~~~~~(12) where
$j$ is the decimal representation of a binary bit string $b_{2N}.......b_{2}b_{1}$
($b_{k}=0$ or $1$). Measurement by Alice using $|\Omega_{j}\rangle's$
leads to an entangled state shared by Bob and Charlie, where Bob possesses
the first particle and Charlie the others `$N$' particles. After
that, Bob has to perform a one-particle measurement using 
\{$\frac{|0\rangle+|1\rangle}{\sqrt{2}},\frac{|0\rangle-|1\rangle}{\sqrt{2}}$\}
and convey his outcome to Charlie. Then Charlie will be able to obtain
the unknown $N$-qubit state, by performing a suitable unitary operation
on his qubits. Hence, QIS of arbitrary $N$-qubit state is achieved
using this channel.

~~

\subsection*{~~~~~~~~~~~~~~~~~~~~~~~~~~~~~~~~~~~~~~~{\normalsize ~V.
CONCLUSION }}

We have demonstrated a number of useful applications in quantum communication
using a general odd particle state which can be obtained by the combination
of one GHZ and many Bell pairs. We have shown that, the five-qubit
state has maximum entanglement between two sub-parties. Arbitrary
one and two qubit states teleportation and quantum information splitting
have been carried out explicitly using the five-qubit state. Three
different protocols are exploited for single qubit QIS. We have also
shown that, teleportation as well as QIS can be generalized using
this state. This state works perfectly for superdense coding and the
capacity of superdense coding reaches the `Holevo bound\textquoteright{}
for `$2N+1$\textquoteright{} classical bits transfer via `$N+1$\textquoteright{}
qubits. The fact that, Bell and GHZ states are realized in laboratory
conditions makes our protocol experimentally achievable. The decoherance
properties and NDD of this state can also be investigated {[}37-40{]}.

~~

~~

\paragraph*{{\large References :}}

~~~~

~~~~

{\small {[}1{]} M. A. Nielsen and I. L. Chuang, `Quantum Computation
and Quantum Information', (Cambridge Univ. Press, 2002).}{\small \par}

{\small {[}2{]} C. H. Bennett, G. Brassard, C. Crepeau, R. Jozsa,
A. Peres, and W. K. Wootters, Phys. Rev. Lett. 70, 1895 (1993).}{\small \par}

{\small {[}3{]} C. H. Bennett, and S. J. Wiesner, Phys. Rev. Lett.
69, 2881 (1992).}{\small \par}

{\small {[}4{]} M. Hillery, V. Buzek, and A. Berthiaume, Phys. Rev.
A 59, 1829 (1999).}{\small \par}

{\small {[}5{]} D. Gottesman, Phys. Rev. A 61, 042311 (2000).}{\small \par}

{\small {[}6{]} H. J. Briegel, and R. Raussendorf, Phys. Rev. Lett.
86, 910 (2001).}{\small \par}

{\small {[}7{]} D. Bouwmeester, J. W. Pan, K. Mattle, M. Eibl, H.
Weinfurter, and A. Zeilinger, Nature (London) 390, 575 (1997).}{\small \par}

{\small {[}8{]} M. Riebe, H. Hffner, C. F. Roos, W. Hnsel, J. Benhelm,
G. P. T. Lancaster, T. W. Krber, C. Becher, F. S. Kaler, }{\small \par}

{\small ~~~~D. F. V. James, and R. Blatt, Nature (London) 429,
734 (2004).}{\small \par}

{\small {[}9{]} M. D. Barrett, J. Chiaverini, T. Schaetz, J. Britton,
W. M. Itano, J. D. Jost, E. Knill, C. Langer, D. Leibfried, }{\small \par}

{\small ~~~~R. Ozeri, and D. J. Wineland, Nature (London) 429,
737 (2004).}{\small \par}

{\small {[}10{]} I. Marcikic, H. de Riedmatten, W. Tittel, H. Zbinden,
and N. Gisin, Nature (London) 421, 509 (2003).}{\small \par}

{\small {[}11{]} R. Ursin, T. Jennewein, M. Aspelmeyer, R. Kaltenbaek,
M. Lindenthal, P. Walther, and A. Zeilinger, Nature (London)}{\small \par}

{\small ~~~~~430, 849 (2004). }{\small \par}

{\small {[}12{]} S. Bandyopadhyay, Phys. Rev. A 62, 012308 (2000).}{\small \par}

{\small {[}13{]} W. Tittel, H. Zbinden, and N. Gisin, Phys. Rev. A
63, 042301 (2001).}{\small \par}

{\small {[}14{]} C. Schmid, P. Trojek, M. Bourennane, C. Kurtsiefer,
M. Zukowski, and Weinfurter, Phys. Rev. Lett. 95, 230505}{\small \par}

{\small ~~~~~~(2005). }{\small \par}

{\small {[}15{]} F. G. Deng, X. H. Li, C. Y. Li, P. Zhou, and H. Y.
Zhou, Phys. Rev. A 72, 044301 (2005). }{\small \par}

{\small {[}16{]} A. Karlsson, M. Bourennane, Phys. Rev. A 58, 4394
(1998). }{\small \par}

{\small {[}17{]} P. Agrawal, and A. Pati, Phys. Rev. A 74, 062320
(2006).}{\small \par}

{\small {[}18{]} V. N. Gorbachev, A. I. Trubilko, and A. A. Rodichkina,
Phys. Lett. A 314, 267 (2003).}{\small \par}

{\small {[}19{]} S. Prasath, S. Muralidharan, P. K. Panigrahi, and
C. Mitra, eprintquant-ph/0905.1233v2.}{\small \par}

{\small {[}20{]} S. Karumanch, S. Jain, P. K. Panigrahi, R. Srikanth,
and S. Muralidharan, eprint quant-ph/0804.4206v5.}{\small \par}

{\small {[}21{]} H. J. Briegel, and R. Raussendorf, Phys. Rev. Lett.
86, 910 (2001).}{\small \par}

{\small {[}22{]} R. Raussendorf, and Hans J. Briegel, Phys. Rev. Lett.
86, 5188 (2001).}{\small \par}

{\small {[}23{]} I. D. K. Brown, S. Stepney, A. Sudbery, and S. L.
Braunstein, J. Phys. A: Math. Gen. 38(5), 1119-1131 (2005).}{\small \par}

{\small {[}24{]} A. Borras, A. R. Plastino, J. Batle, C. Zander, M.
Casas, and A. Plastino, J. Phys. A: Math. Gen. 40(44), 13407 }{\small \par}

{\small ~~~~~~(2007).}{\small \par}

{\small {[}25{]} S. Muralidharan, and P. K. Panigrahi, Phys. Rev.
A 77, 032321 (2006).}{\small \par}

{\small {[}26{]} S. Choudhury,} {\small S. Muralidharan, and P. K.
Panigrahi, J. Phys. A: Math. Theor. 42, 115303 (2009).}{\small \par}

{\small {[}27{]} G. Rigolin, Phys. Rev. A 71, 032303 (2005).}{\small \par}

{\small {[}28{]} Xiu-Lao Tian, and Xiao-Qiang Xi, eprint quant-ph/0702150v2.}{\small \par}

{\small {[}29{]} A. Pathak, and A. Banerjee, eprint quant-ph/1006.1042v1}.

{\small {[}30{]} Y. Y. Nie, Y. H. Li, J. C. Liu, and M. H. Sang, QINP-273R1
(2011).}{\small \par}

{\small {[}31{]} S. Muralidharan, and P. K. Panigrahi, Phys. Rev.
A 76(6), 062333 (2008).}{\small \par}

{\small {[}32{]} S. Muralidharan, S. Jain, and P. K. Panigrahi, Opt.
Commun. 284 (2011).}{\small \par}

{\small {[}33{]} D. D. B. Rao, P. K. Panigrahi, and C. Mitra, Phys.
Rev. A 78, 022336 (2008). }{\small \par}

{\small {[}34{]} L. Li, and D. Qiu, J. Phys. A: Math. Gen. 40, 10871
(2007). }{\small \par}

{\small {[}35{]} D. Bruss, G. M. D' Ariano, M. Lewenstein, C. Macchiavello,
A. Sen, and U. Sen, Phys. Rev. Lett. 93, 210501 (2004). }{\small \par}

{\small {[}36{]}} {\small S. Muralidharan, S. Karumanchi, S. Narayanaswamy.
R. Srikanth, and P. K. Panigrahi, eprint quant-ph/0907.3532v2.}{\small \par}

{\small {[}37{]} P. K. Panigrahi, M. Gupta, A. Pathak, and R. Srikanth,
AIP, 864, 197 (2006).}{\small \par}

{\small {[}38{]} M. Gupta, A. Pathak, R. Srikanth, and P. K. Panigrahi,
IJQI, 5, 62 (2007).}{\small \par}

{\small {[}39{]} J. R. Samal, M. Gupta, P. K. Panigrahi, and A. Kumar,
J. Phys. B: At. Mol. Opt. Phys. 43, 095508 (2010). }{\small \par}

{\small {[}40{]} S. Jain, P. K. Panigrahi, and S. Muralidharan, EPL,
87, 6008 (2009).}
\end{document}